\documentclass{article}
\usepackage{emulateapj,psfig}
\usepackage{graphicx}
\usepackage{timesfonts}
\def\arraystretch{1.25}

\makeatletter
\newenvironment{inlinetable}{%
\def\@captype{table}%
\noindent\begin{minipage}{0.999\linewidth}\begin{center}}
{\end{center}\end{minipage}\smallskip}

\newenvironment{inlinefigure}{%
\def\@captype{figure}%
\noindent\begin{minipage}{0.999\linewidth}\begin{center}}
{\end{center}\end{minipage}\smallskip}
\makeatother
\def\ltsima{$\; \buildrel < \over \sim \;$}
\def\lsim{\lower.5ex\hbox{\ltsima}}
\def\loe{\lower.5ex\hbox{\ltsima}}
\def\gtsima{$\; \buildrel > \over \sim \;$}
\def\gsim{\lower.5ex\hbox{\gtsima}}
\def\goe{\lower.5ex\hbox{\gtsima}}

\newcommand{\cdo}{$^{12}C(\alpha ,\gamma)^{16}O\;$}
\newcommand{\dydz}{$\Delta Y / \Delta Z\;$}

\lefthead{Zoccali et al.}

\righthead{On the Helium content of GGCs}

\begin{document}

\title{On the helium content of Galactic globular clusters via the R parameter
\footnote{Based on observations with the NASA/ESA {\it Hubble Space Telescope},
obtained at the Space Telescope Science Institute, which is operated
by AURA, Inc., under NASA contract NAS5-26555, and on observations
retrieved from the ESO ST-ECF Archive.}  }

\author{
M. Zoccali\altaffilmark{1}, 
S. Cassisi\altaffilmark{2,3}, 
G. Bono\altaffilmark{4},
G. Piotto\altaffilmark{1},
R. M. Rich\altaffilmark{5}
S. G. Djorgovski\altaffilmark{6} 
}

\affil{1. Dipartimento di Astronomia, Universit\`a di Padova,
vicolo dell'Osservatorio 5, 35122 Padova; zoccali@pd.astro.it,
piotto@pd.astro.it}
\affil{2. Osservatorio Astronomico di Collurania, via M. Maggini,
64100 Teramo; cassisi@astrte.te.astro.it}
\affil{3. Max-Planck-Institut fur Astrophysik, Karl-Schwarzschild-Str. 1, 
85740 Garching  b. Munich , Germany}
\affil{4. Osservatorio Astronomico di Roma, Via Frascati 33,
00040 Monte Porzio Catone, Italy; bono@coma.mporzio.astro.it}
\affil{5. Department of Physics and Astronomy, Division of Astronomy
and Astrophysics, University of California, Los Angeles, CA 90095-1562;
rmr@astro.ucla.edu}
\affil{6. California Institute of Technology, MS 105-24,
Pasadena, CA 91125; george@astro.caltech.edu}

\begin{abstract}

We estimate the empirical R parameter in 26 Galactic Globular Clusters
(GGCs) covering a wide metallicity range, imaged by WFPC2 on board the
Hubble Space Telescope.  The improved spatial resolution permits a
large fraction of the evolved stars to be measured and permits
accurate assessment of radial populaton gradients and completeness
corrections. In order to evaluate both the He abundance and the He to
metal enrichment ratio, we construct a large set of evolutionary
models by adopting similar metallicities and different He contents. We
find an absolute He abundance which is lower than that estimated from
spectroscopic measurements in HII regions and from primordial
nucleosynthesis models.  This discrepancy could be removed by adopting
a \cdo nuclear cross section about a factor of two smaller than the
canonical value, although also different assumptions for mixing
processes can introduce systematical effects.

The trend in the R parameter toward solar metallicity is consistent
with an upper limit to the He to metal enrichment ratio of the order
of 2.5. Detailed calculations of central He-burning times as a
function of the HB morphology suggest that He lifetimes for hot HB
stars are on average $\approx$ 20\% longer than for RR Lyrae and red
HB stars. Therefore, the increase in the empirical R values of
metal-poor clusters characterized by blue HB morphologies is due to an
increase in the HB lifetime and not due to an increase in the He
abundance.

\end{abstract}

\keywords{globular clusters: general -- stars: abundances -- 
stars: evolution -- stars: horizontal branch}  

\section{Introduction} 

The He abundance is fundamental in several astrophysical problems.
Big bang nucleosynthesis models supply tight predictions on the
primordial He content, and therefore empirical estimates of this
parameter are crucial for constraining their plausibility (Hogan,
Olive, \& Scully 1997).  At the same time, stellar evolutionary and
pulsational models do require the assumption of a He to metal
enrichment ratio \dydz in order to reproduce the observed properties
of both metal-poor and metal-rich stellar structures (Bono et
al. 1997a). Observational constraints on this parameter can improve
the accuracy of several theoretical observables, and in particular of
stellar yields predicted by Galactic chemical evolution models
(Tsujimoto et al. 1997; Pagel \& Portinari 1998).

One of the most widely used methods for {\em measuring} the He
abundance is to measure fluxes of nebular emission lines in planetary
nebulae (Peimbert 1995) or in extragalactic H~II regions (Pagel et
al. 1992; Izotov, Thuan, \& Lipovetsky 1997; Olive, Steigman, \&
Skillman 1997).  Independent estimates based on high signal-to-noise
measurements of He abundance give very similar results
($Y=0.23-0.24$), thus suggesting that the empirical uncertainties are
quite small. However, by adopting detailed radiative transfer
calculations of H and He, Sasselov \& Goldwirth (1995) supported the
evidence that current He measurements could be affected by large
systematic errors. The He content can also be obtained by direct
spectroscopic measurements in hot Horizontal Branch (HB)
stars. Unfortunately, these stars are affected by gravitational
settling and by radiation levitation (Michaud et al. 1983; Moehler et
al. 1999), and therefore they might present peculiar abundance
patterns.  Nevertheless, there seems to be a consensus that the
primordial He abundance should not be lower than Y=0.22 (Olive,
Steigman, \& Walker 1999).

Both absolute and/or relative He abundances can also be {\em
estimated} because the evolution of population II stars is sensitive
to the primordial helium abundance.  The first helium-sensitive
indicator to be identified was the R parameter, defined as the ratio
between the number of stars along the HB and the number of red giant
branch (RGB) stars brighter than the HB luminosity
($R=N_{HB}/N_{RGB}$, Iben 1968).  Additional parameters also use the
helium burning stars of the horizontal branch as abundance indicators.
The $\Delta$ parameter is the magnitude difference between HB stars
and main sequence (MS) stars (Carney 1980) and the $A$ parameter is
the mass-luminosity exponent of RR Lyrae stars (Caputo, Cayrel, \&
Cayrel de Strobel 1983).  The fine structure of the main sequence
locus of population II stars (Faulkner 1967) is also a potentially
powerful method to constrain \dydz.  On the basis of Hipparcos
parallaxes Pagel \& Portinari (1998) investigated the fine structure
of solar neighborhood MS stars and found that the current estimates of
\dydz are still affected by large uncertainties. The other three
methods have been recently applied by Sandquist (1999, hereinafter
S99) to a sample of 42 Galactic Globular Clusters (GGCs).  Sandquist
also comprehensively discusses the pros and cons of these abundance
indicators and in particular the statistical and systematic errors
affecting both absolute and relative He estimates.  The results by S99
support the evidence that both the $\Delta$ and the $A$ parameter can
only give reliable relative He abundances, due to current
uncertainties on the metallicity scale and on the RR Lyrae temperature
scale.  At the same time, S99 brought out that absolute He abundances
based on the R parameter ($Y\approx0.2$) could also be affected by
additional systematic errors, and that both relative and absolute
estimates do not show, within current uncertainties, a clear evidence
of a trend with metallicity. The latter finding does not support the
results by Renzini (1994), Minniti (1995), Bertelli et al. (1996), and
Desidera, Bertelli, \& Ortolani (1998) who suggest that in the
Galactic bulge the He abundance scales with metallicity according to a
slope ranging from 2 to 3.5.  Moreover detailed comparisons between
solar standard models and accurate helioseismic data (Ciacio,
Degl'Innocenti \& Ricci 1997; Degl'Innocenti et al. 1997;
Christensen-Dalsgaard 1998) is more consistent with
\dydz$\approx2$. The large spread in the empirical values suggests
that current He estimates are still hampered by large uncertainties
which do not allow us to disentangle the intrinsic variation, if any,
from systematic effects.
 
The empirical evaluation of the R parameter relies only on star
counts. Nevertheless, misleading effects can be introduced by the
method adopted for fixing the Zero Age Horizontal Branch (ZAHB)
luminosity, by differential reddening, as well as by the occurrence of
population gradients inside the cluster (Buzzoni et al. 1983; Caputo,
Martinez Roger, \& Paez 1987; Djorgovski \& Piotto 1993; Bono et
al. 1995; S99).  The He abundance is estimated by comparing observed
values with the ratio of HB and RGB evolutionary times, which relies
on evolutionary predictions characterized by a negligible dependence
on stellar age (Iben \& Rood 1969). The He burning lifetimes do depend
on input physics such as equation of state, opacity, and nuclear cross
sections (Brocato, Castellani, \& Villante 1998; Cassisi et al. 1998)
adopted to construct HB models as well as on the algorithm adopted for
treating the mixing processes (Sweigart 1990, and references therein).

The main aim of this investigation is to derive the R parameter for a
sample of 26 GGCs, and to compare empirical values with theoretical
predictions in order to gather information on both the He content and
its trend with metallicity. In order to accomplish this goal we
specifically calculate a large set of HB models, adopting the most
up-to-date input physics.  We rely on the high number of stars sampled
in each cluster, on the wide metallicity range covered by clusters,
and on the high homogeneity of theoretical predictions and data to
constrain the behavior of both R and \dydz parameters.

\section{The cluster database}

We evaluate the R parameter in 26 GGCs of our HST database (Zoccali et
al. 1999, hereinafter Z99, and references therein) including images
from the HST projects GO-6095 and GO-7470, and similar data retrieved
from the HST archive.  To avoid systematic uncertainties in the star
counts, we exclude from the sample the clusters affected by strong
foreground contamination (NGC~6522, NGC~6441), as well as those having
hot HB stars close to the magnitude limit and for which the
completeness correction was not estimated (NGC~6205).

The R parameter is defined as $N_{HB}/N_{RGB}$, where $N_{RGB}$ is the
number of RGB stars brighter than a reference luminosity, generally
fixed according to the luminosity of RR Lyrae stars (Buzzoni et
al. 1983) or to the ZAHB luminosity (Bono et al. 1995).  However, the
accurate determination of this luminosity is often a thorny problem.
In fact, together with the well-known difficulties in estimating the
RR Lyrae luminosity and/or the ZAHB luminosity for clusters with only
red or blue HB morphologies, we are also dealing with the problem of
the differential bolometric correction between HB and RGB stars.  This
means that, as soon as the HB luminosity/magnitude is fixed, the
proper RGB magnitude at this luminosity can be estimated only by
accounting for the change in the bolometric correction between HB and
RGB structures. Current estimates have been derived by adopting
different assumptions on the value of the bolometric correction and on
its variation with metallicity. Therefore, the comparison among the R
values available in the literature it is not straightforward even for
the same clusters.

\begin{inlinetable}
\begin{center}
\tablewidth{0pt}
\small
\begin{tabular}{cccrrcc}
\multicolumn{7}{c}{TABLE 1. Cluster Data}\\
\hline
NGC & [M/H] & $V^a_{\rm ZAHB}$ & $N_{\rm HB}^c$ & $N_{\rm RGB}^d$ &
R$^e$ & $N_{\rm B}/N_{\rm R}^f$\\
\hline
~104 & --0.54 & 14.26     & ~358 & 235 &  1.52$\pm$0.13  & 0/358   \nl  
~362 & --0.84 & 15.66$^b$ & ~247 & 197 &  1.25$\pm$0.12  & 17/215  \nl  
1851 & --0.93 & 16.33$^b$ & ~297 & 237 &  1.24$\pm$0.11  & 89/180  \nl  
1904 & --1.19 & 16.31     & ~177 & 116 &  1.53$\pm$0.18  & 0/172   \nl  
2808 & --1.03 & 16.50     & ~851 & 606 &  1.26$\pm$0.06  & 462/389 \nl  
4590 & --1.68 & 15.72     & ~~34 & ~35 &  0.93$\pm$0.23  & 23/11   \nl     
5634 & --1.45 & 18.04     & ~146 & 105 &  1.33$\pm$0.17  & 131/2   \nl  
5694 & --1.49 & 18.73     & ~249 & 159 &  1.50$\pm$0.15  & 241/5   \nl  
5824 & --1.48 & 18.55     & ~520 & 372 &  1.46$\pm$0.10  & 480/20  \nl  
5927 & --0.17 & 16.94     & ~195 & 136 &  1.47$\pm$0.17  & 0/195   \nl 
5946 & --1.24 & 17.45     & ~113 & ~96 &  1.13$\pm$0.11  & 113/0   \nl  
5986 & --1.31 & 16.54     & ~237 & 154 &  1.48$\pm$0.15  & 224/4   \nl  
6093 & --1.27 & 16.46     & ~263 & 221 &  1.25$\pm$0.11  & 251/12  \nl  
6139 & --1.29 & 18.50     & ~299 & 223 &  1.28$\pm$0.11  & 290/9   \nl  
6235 & --1.06 & 17.42$^b$ & ~~35 & ~37 &  0.90$\pm$0.22  & 31/2    \nl  
6284 & --0.99 & 17.36$^b$ & ~132 & 104 &  1.33$\pm$0.17  & 132/0   \nl  
6287 & --1.67 & 17.29     & ~~92 & ~59 &  1.49$\pm$0.25  & 82/9    \nl  
6293 & --1.55 & 16.56     & ~137 & 101 &  1.30$\pm$0.17  & 130/4   \nl  
6342 & --0.43 & 17.66     & ~~73 & ~42 &  1.74$\pm$0.34  & 0/73    \nl  
6356 & --0.30 & 18.15     & ~370 & 231 &  1.60$\pm$0.13  & 0/370   \nl  
6362 & --0.82 & 15.63$^b$ & ~~38 & ~33 &  1.15$\pm$0.27  & 6/27    \nl  
6388 & --0.39 & 17.41     & 1353 & 747 &  1.74$\pm$0.07  & 202/1151  \nl  
6624 & --0.22 & 16.30     & ~123 & ~86 &  1.43$\pm$0.20  & 0/123  \nl  
6652 & --0.72 & 16.21     & ~~62 & ~47 &  1.32$\pm$0.26  & 0/62    \nl  
6981 & --1.19 & 17.32     & ~~65 & ~56 &  1.16$\pm$0.21  & 21/24   \nl  
7078 & --1.82 & 15.93     & ~390 & 242 &  1.57$\pm$0.13  & 292/50 \nl 
\hline
\end{tabular}
\end{center}
\normalsize
\begin{minipage}{1.00\linewidth}
\noindent{$^a$ For the $V_{ZAHB}$ determination see the 
discussion in Z99. $^b$ Due to a misidentification of some blue 
RR Lyrae stars, the $V_{ZAHB}$ of these clusters were underestimated 
by few hundredth of magnitude in Z99. $^{c,d}$ The total
number of HB and RGB stars, respectively. $^e$ The error budget on R 
includes: the Poisson error on the raw star counts, the completeness 
correction, and the weighted mean of the three radial determinations. 
$^f$ Number of stars bluer/redder than the RR Lyrae gap.} 
\end{minipage}
\end{inlinetable}

To overcome systematic errors introduced both by metallicity and
gravity variations, in the present analysis we choose to ``save the
observables'', i.e. we define $N_{RGB}$ as the number of RGB stars
brighter than the ZAHB $V$ magnitude ($V \le V_{ZAHB}$).  As a
consequence, both $t_{HB}$ and $t_{RGB}$ values are estimated after
theoretical predictions are transformed to the observational
plane. Table~1 lists the cluster name, its global metallicity, and the
other observed quantities.  Owing to calibration problems with WFPC2
images (Stetson 1998, 1999), the values of $V_{ZAHB}$ adopted for this
work could be affected by an uncertainty $\le 0.1$ mag, and therefore
the values in Table~1 have to be considered only as {\it relative}
evaluations.  Cluster metallicities are based on the Cohen et
al. (1999) metallicity scale, while global metallicities were
estimated by assuming a mean $\alpha$-element enhancement of
[$\alpha$/Fe]=$0.3$ for [Fe/H]$\le-1$, and [$\alpha$/Fe]=$0.2$ for
[Fe/H]$>-1$ (see Z99).
   
The observed star counts have been corrected for completeness.  Since
crowding effects depend on the distance from the cluster center, we
divide each field in three radial annuli, and then correct the counts
in each annulus by using the completeness correction appropriate for
each magnitude level. As a consequence, we compute three independent
radial values of R, and our final R value is their weighted mean.
This approach also gives a check for spurious radial trends which
could be caused either by population gradients (Djorgovski \& Piotto
1993) or by an overestimate (underestimate) of the completeness
correction.  Interestingly enough, two clusters (NGC~6273 and
NGC~6934) show a strong variation of R with the distance from the
cluster center.  Due to the small area covered by the WFPC2 field, and
to the different pixel size --- hence different sampling --- of the
most central chip (PC) when compared with to the outer ones (WFs) we
cannot firmly assess whether this behavior is intrinsic ---
i.e. caused by a radial population gradient --- or caused by
systematic errors in the completeness correction.  Radial color
gradients in NGC~6934 have been already found by Sohn,
Byun \& Chun (1996), we are not aware of wide field investigations
on NGC~6273.  The peculiar behavior of star counts in
these clusters deserves a detailed investigation,
therefore we exclude both of them from the sample.

The completeness correction for red HB stars is assumed to be
identical to that for RGB stars at the same magnitude. However,
completeness correction for blue HB stars could be significantly
different from those derived for the RGB stars. In fact, at fixed $V$
magnitude, blue HB stars have brighter $B$ magnitudes, and thus a
higher probability to be detected. For some very blue HB clusters,
namely NGC~2808, NGC~6273, and NGC~7078, we perform several direct
experiments by adding to the frames artificial stars along the HB
fiducial line. We find that the completeness correction for blue HB
stars is a factor of 1.046 higher than the completeness for RGB and MS
stars located at the same $V$ magnitude. This difference appears
fairly constant in different clusters and over a large magnitude
range. As a consequence, the artificial-star tests for the other
clusters are performed only for RGB/MS stars, and the completeness
correction for HB stars was scaled according to the same factor of
1.046. It is worth noting that such a completeness correction is
significant only for extreme blue HB stars and that the correction to
the R parameter is always smaller than 5\%.  The only exception is
NGC~2808, a cluster which shows a very long and populated HB blue
tail, and a very high central density. By applying the completeness
correction to this cluster the R parameter changed by $11\%$.

\section{Discussion}

Figure 1 shows the comparison between empirical R values and theoretical
predictions at fixed age (14 Gyr) for three different assumptions
about the He content and for metal abundances ranging from
[M/H]=$-2.2$ to solar chemical composition. At fixed composition, a
large set of evolutionary models for both H and He burning phases was
constructed by adopting the input physics already discussed by Cassisi
\& Salaris (1997, hereinafter CS). As usual, the HB lifetime is 
estimated on the basis of the HB model located at $\log T_e=3.85$,
i.e. by assuming as representative of $t_{HB}$ the central He burning
time of a structure whose ZAHB is inside the RR Lyrae instability
strip. Our calculations confirm Iben's (1968) original finding
that the R parameter is virtually independent of the adopted cluster
age.  The comparison between theory and observations also confirms the
finding by S99 that the absolute He content resulting from the
R-method is $Y\approx 0.20$.  As discussed in Section 1, this value is
significantly lower than the canonical He abundance expected from the
primordial nucleosynthesis and measured in the HII regions, showing
that there is some systematic uncertainty affecting the calibration of
R as a function of Y.

\begin{inlinefigure}
\centerline{\includegraphics
[height=0.62\linewidth, width=1\linewidth]{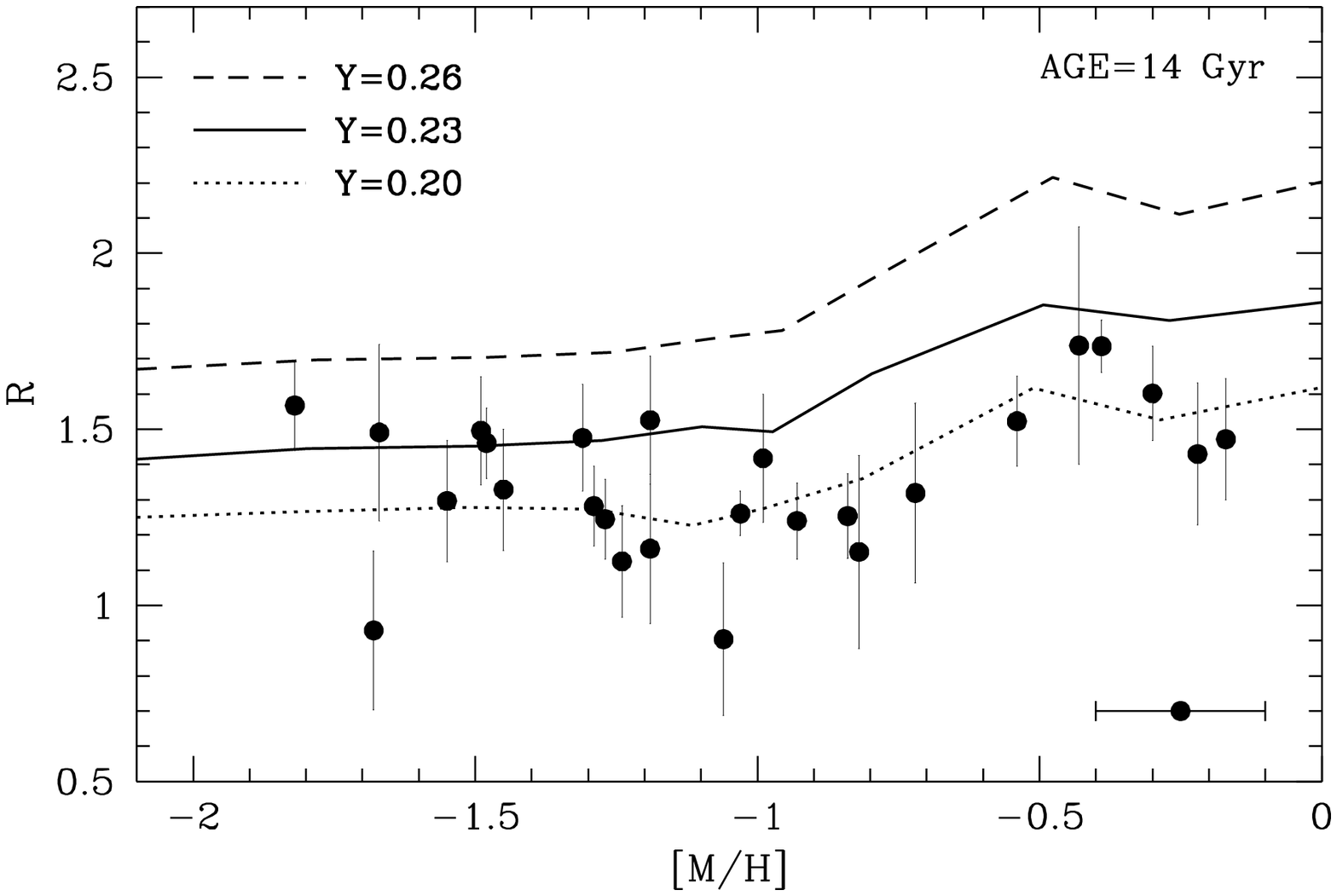}}
\caption{R parameter vs. global metallicity. Theoretical predictions 
are estimated at fixed age (14 Gyr) and for three different initial 
He abundances. The horizontal error bar accounts for current 
uncertainties on metallicity.}
\label{fig1}
\end{inlinefigure}

\begin{inlinefigure}
\centerline{\includegraphics
[height=0.62\linewidth, width=1\linewidth]{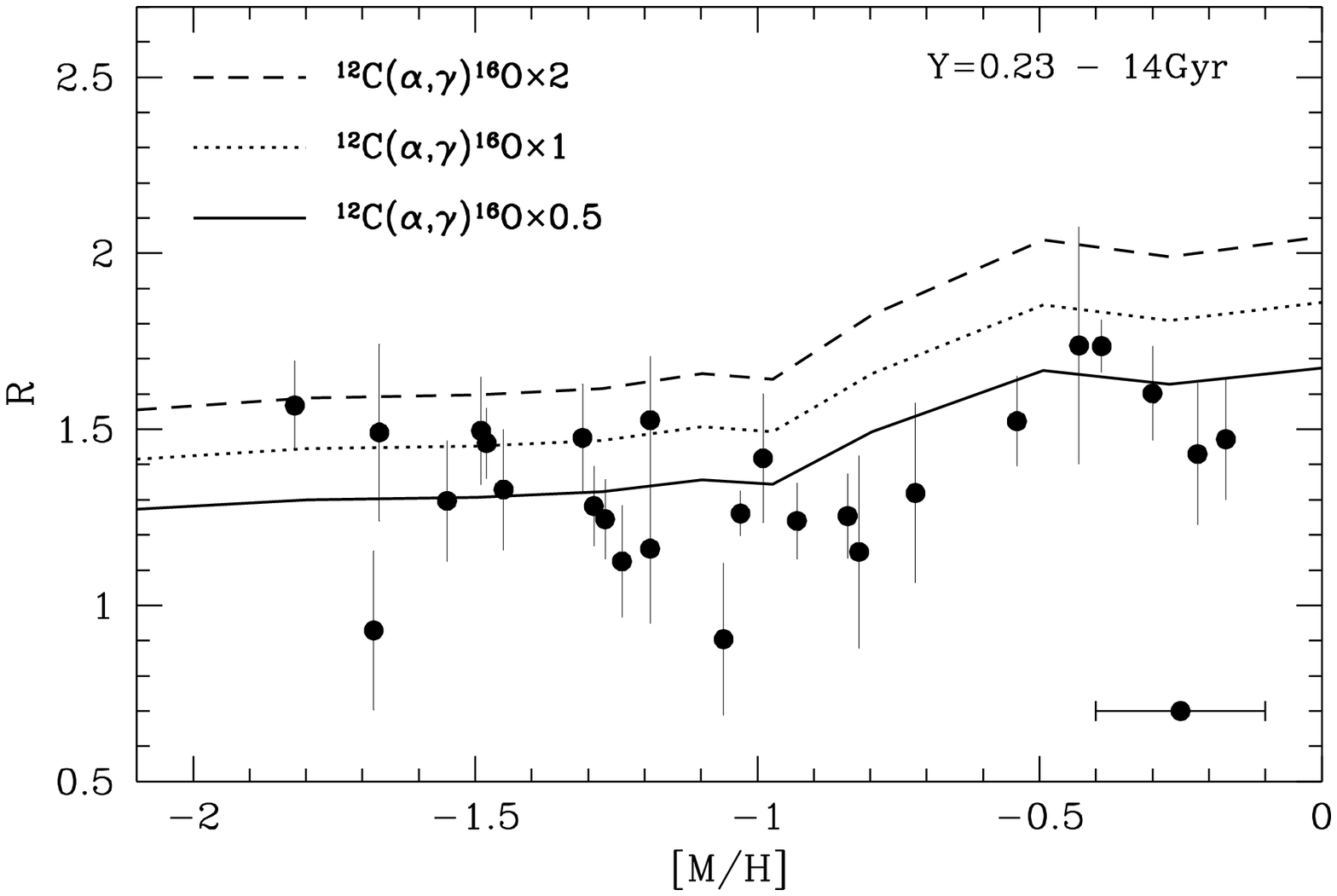}}
\caption{Same as Fig. 1, but theoretical predictions refer to HB 
models at fixed age and He content constructed by assuming an 
increase/decrease of a factor of two in the \cdo nuclear cross 
section.}
\label{fig2}
\end{inlinefigure}

As discussed in Brocato et al. (1998) and Cassisi et al. (1998), among
the physical input parameters that govern the HB lifetime, the nuclear
cross section for the \cdo reaction is affected by the largest
uncertainty.  In order to investigate the dependence of $t_{HB}$ on
the poorly measured \cdo cross section, we performed several numerical
experiments.  Figure 2 displays the dependence of the theoretical R
values on an increase/decrease by a factor of two in the efficiency of
the quoted reaction when compared with the value provided by Caughlan
et al.  (1985). Even though the range of the nuclear cross section
values used in Fig. 2 is quite large, it is still inside the current
error of its empirical measurements (Buchmann 1996). This fact clearly
shows the sensitivity of the predicted R values on input physics, and
suggests that the R method can not be presently used for the
determination of the absolute He abundance. Instead, we can use the R
parameter to constrain the input parameters of the model. If we rely
on primordial He abundance measurements in extragalactic HII regions,
and we assume that mixing processes have been properly accounted for
in current HB models, the data plotted in Fig. 2 suggest that the
current value of the \cdo nuclear cross section should be roughly a
factor of two smaller.

\begin{inlinefigure}
\centerline{\includegraphics
[height=0.62\linewidth, width=1\linewidth]{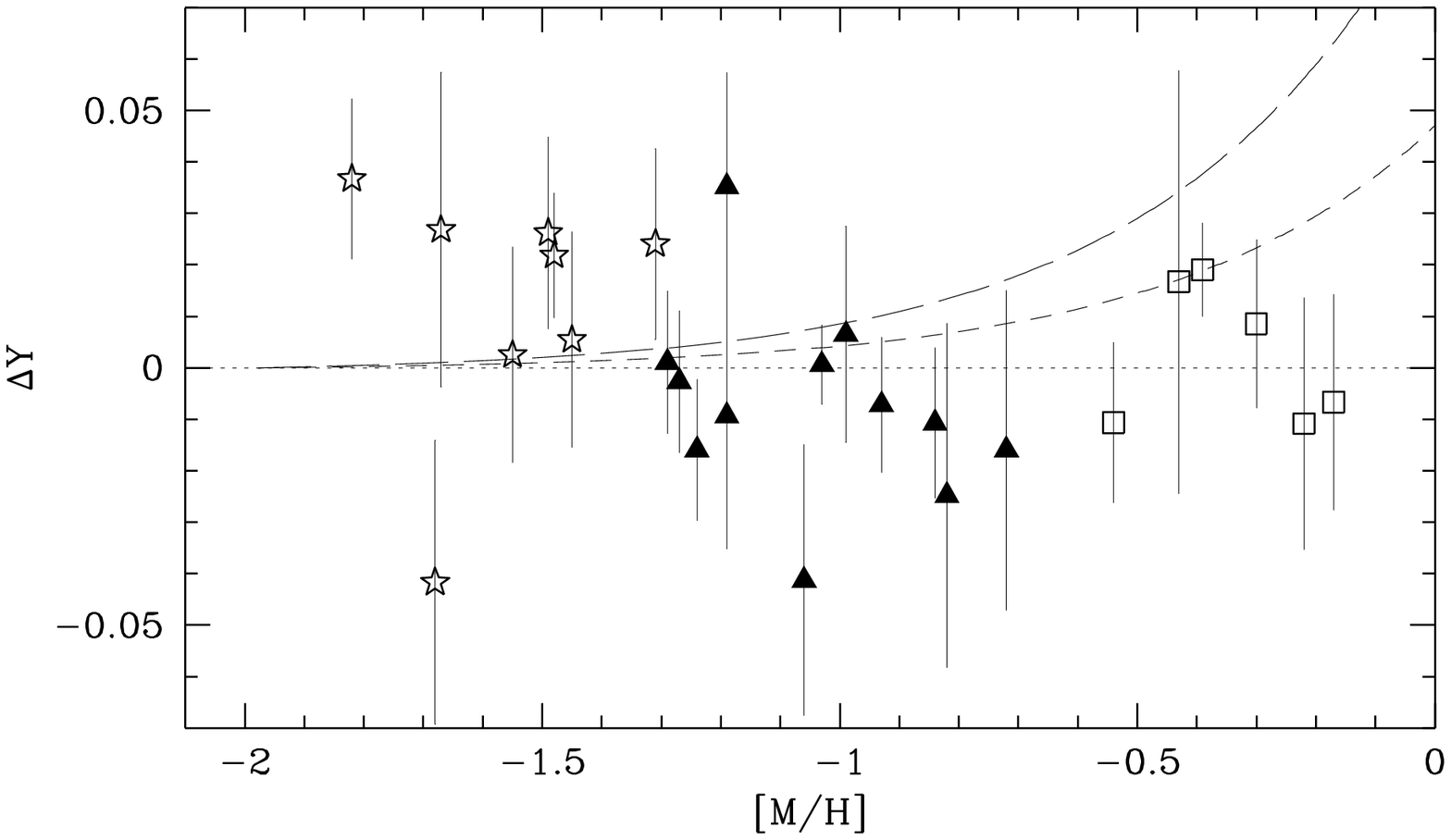}}
\caption{Relative He abundance vs. global metallicity.  The residuals 
are obtained by differencing the R values from the theoretical line
for Y=0.20 in Fig. 1.  The different symbols for the cluster correspond
to low, intermediate, and high metallicity.  Short-dashed line gives
the helium to metal enrichment ratio using \dydz$=2.5$ while the upper 
long-dashed line (which clearly misses the metal rich clusters) is for 
\dydz$=5$.}
\label{fig3}
\end{inlinefigure}

This notwithstanding, we can still use the R parameter to constrain
the still controversial \dydz value. First of all, we note that even
for constant Y, the R values plotted in Fig. 1 present a trend with
metallicity.  For [M/H]$> -1$ the empirical R values show an increase
and then a flat distribution.  As already noted by Desidera et al.
(1997), this behavior is due to the fact that an increase in
metallicity causes the luminosity of the RGB bump to become fainter
than the ZAHB luminosity, as shown by Z99, and it is well reproduced
by the models (Fig. 4).  In order to investigate the trend of the
relative He content with the global metallicity, in Fig. 3 we plot the
residuals of the measured R values with respect to the model that
better reproduces the data in Fig. 1, i.e. Y=0.20.  We emphasize that
the absolute value of Y adopted as reference only changes the vertical
zero point of Fig. 3.  We would obtain the same distribution using a
Y=0.23 model with a \cdo smaller by a factor of two.  The two dashed
lines show the expected variation in the He abundance using two
different assumed values for \dydz .  The present data suggest that
the upper limit to the He to metal enrichment ratio should be
$\approx$2.5.
In particular the data would appear to exclude \dydz$\sim 5$ because
this would require to count, at metallicity [M/H]$\sim-0.3$, a number
of HB stars lower than that observed by about $30\%$, or, conversely,
a number of RGB stars higher by the same amount.  Although toward
higher metallicity our counts could be contaminated by upper AGB
stars, the stellar lifetimes of this objects can not account for such
a high effect.. On the other hand, the number of red HB stars can be
affected by systematic error, although not as high as $30\%$, only for
the clusters affected by strong differential reddening, such as
NGC~6388 (because the HB red clump merges into the RGB).  The CMD of
other metal-rich clusters (NGC~5927 and NGC~6624) show very well
separated HB and RGB (Sosin et al.\ 1997) and can not be affected by
such a problem.
Our finding of a \dydz$\le 2.5$ confirms the original results by Peimbert
\& Torres-Peimbert (1976), and more recently by Peimbert \& Peimbert
(2000, and references therein).

\begin{inlinefigure}
\centerline{\includegraphics
[height=1\linewidth, width=1\linewidth]{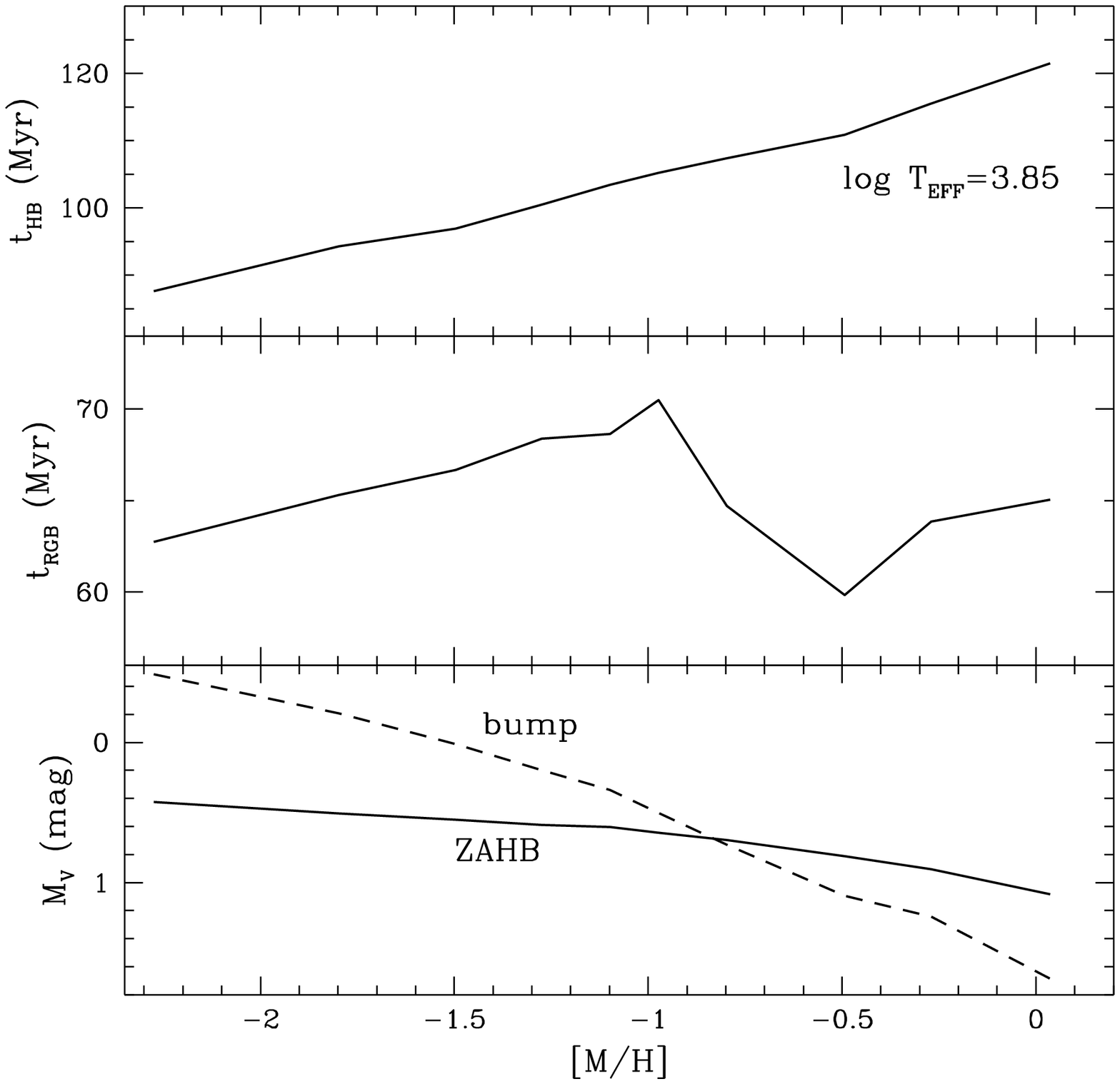}}
\caption{Top: HB evolutionary time vs. global metallicity at 
fixed age (14 Gyr) and He content (0.23) estimated at the RR Lyrae 
instability strip ($\log Te=3.85$). The dashed line displays selected 
$t_{HB}$ values for hot HB stars ($\log Te=4.48$). Middle: RGB evolutionary 
time. Bottom: ZAHB (solid line) and RGB bump (dashed line) visual 
magnitude.}
\label{fig4}
\end{inlinefigure}

Figure~3 shows another interesting feature: metal-poor GGCs present on
average a larger R value.  A similar conclusion was recently reached
in S99, who also note that GGCs with blue HB morphology also have
systematically large R parameters; they offer no explanation for this
effect.
Keeping in mind this problem, we investigate the behavior of each
single term in the predicted time ratio. Figure 4 shows the key
theoretical ingredients of this parameter i.e. $t_{HB}$ (top),
$t_{RGB}$ (middle), and the magnitudes of both ZAHB and RGB bump
(bottom) as a function of metallicity. The dependence of both
$M_V(ZAHB)$ and $M_V(bump)$ on metallicity accounts for the slope
disclosed by both empirical and theoretical R values (see Fig. 1). In
fact, for metallicity larger than [M/H]$\approx-1$, the $t_{RGB}$
presents a sudden decrease caused by the fact that the RGB bump
becomes fainter than the ZAHB. This occurrence implies a strong
decrease in $t_{RGB}$, since, for clusters in this metallicity range,
the RGB bump phase alone contributes for $\approx$20\% to the total
$t_{RGB}$ value.  This means that small changes in $M_V(ZAHB)$ can
cause substantial variations in the R value. This effect explains why
the R values of the intermediate metallicity clusters present a large
scatter.

The values of $t_{HB}$ shown in Fig.~4 refer to a star located in the
middle of the RR Lyrae instability strip ($\log T_e=3.85$).  However,
current evolutionary calculations (Castellani et al. 1994) suggest
that the lifetime of blue tail HB stars is roughly 30\% longer when
compared with the lifetimes of HB stars located inside the instability
strip. As a consequence, to assess whether predicted R values are
affected by systematic uncertainties we must explore in more detail
the dependence of the R parameter on $t_{HB}$ and in particular on HB
morphology.  We split the HB into three different regions, namely the
HB stars bluer than RR Lyrae variables (B), the RR Lyrae variables
(V), and the HB stars redder than RR Lyrae variables (R) (Lee,
Demarque, \& Zinn 1994).  As a plausible estimate for the blue and the
red edge of the RR Lyrae instability strip we adopted, according to
Bono et al. (1997b), $T_e\approx7300$ and 5900 K respectively. On the
basis of these ingredients, and by assuming a linear mass distribution
along the HB, we estimated the average HB lifetime,
$\overline{t_{HB}}$, at fixed He abundance $Y=0.20$, for stars
belonging to B, V, and R regions respectively.  Figure 5 shows the
theoretical R parameters for the three different HB morphologies.  The
R values based on the HB lifetime of RR Lyrae stars ---
$\overline{t_{HB}}(V)$ --- are almost identical to those based on
$\overline{t_{HB}}(R)$. Thus suggesting that $t_{HB}(\log T_e=3.85)$
is representative of central He-burning lifetime for GCs characterized
by HB stars with $\log T_e\le 3.86$.  On the other hand, we find that
the lifetimes of blue HB stars --- $\overline{t_{HB}}(B)$ --- are
approximately 20\% longer than $\overline{t_{HB}}(V)$ and
$\overline{t_{HB}}(R)$. This means that the R values of GGCs with blue
HB morphologies are expected to be $\approx 0.25$ units higher than
those with red HBs. Therefore the observed high R values in metal-poor
clusters are not due to a real increase in the He abundance, but are
likely the consequence of their blue HB morphology. The large scatter
among metal-poor clusters is mainly due to the different stellar
distributions along the blue tail (see Fig.\ 9 in Piotto et al. 1999).

\begin{inlinefigure}
\centerline{\includegraphics 
[height=0.62\linewidth, width=1\linewidth]{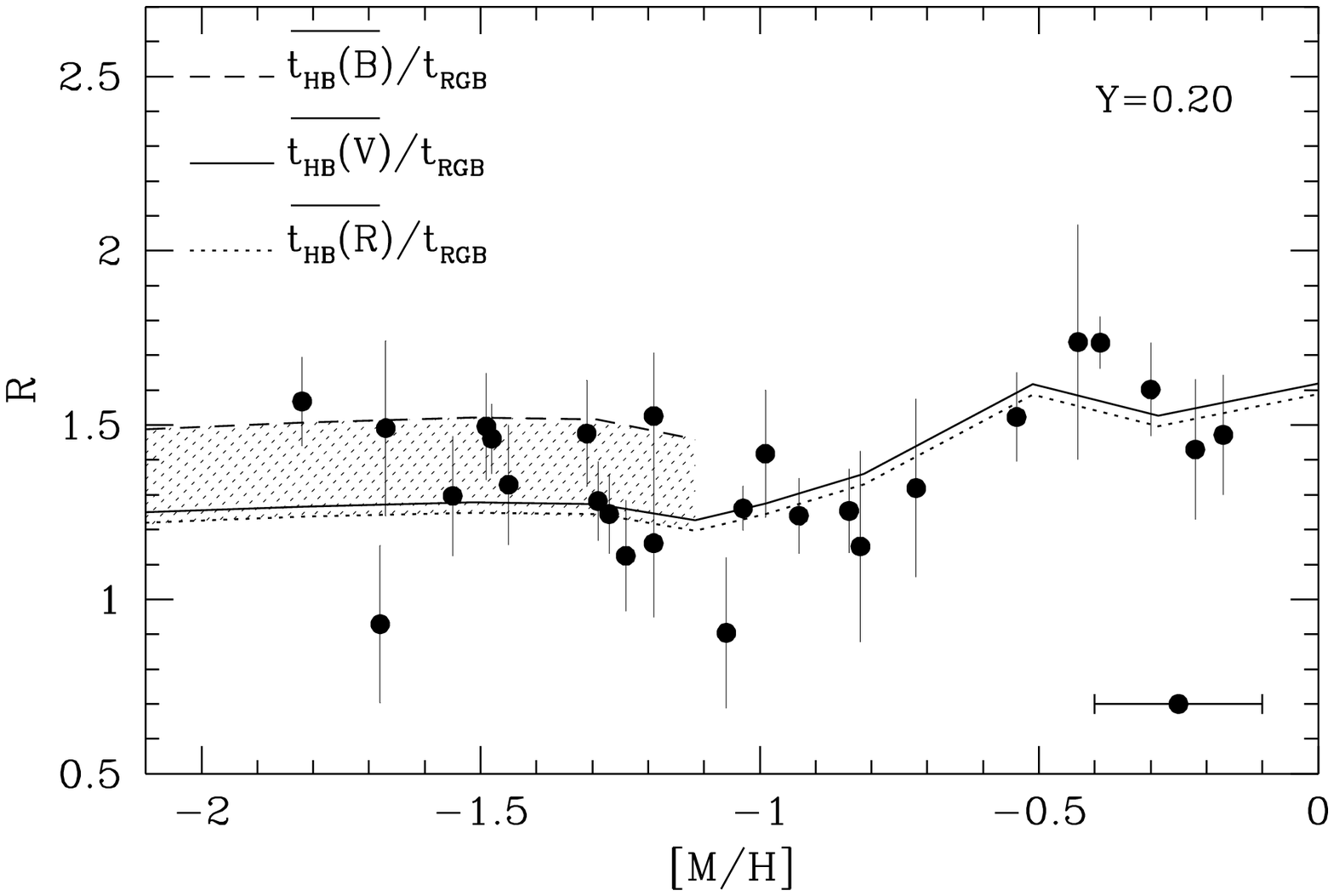}}
\caption{Comparison between predicted and empirical R values, showing
that changes in the HB lifetime as a function of metallicity is the
dominant variable affecting the trend of R with metallicity. 
Theoretical R values are estimated by adopting a fixed He content and 
an average HB lifetime which accounts for three different HB morphologies. 
{\rm B} refers to HB stars bluer than RR Lyrae variables (dashed line), 
{\rm V} to HB stars located inside the instability strip (solid line), 
and {\rm R} to HB stars redder than RR Lyrae variables (dotted line).
The hatched area shows the region within which GGCs characterized by a 
blue HB morphology are expected to be located.}
\label{fig5}
\end{inlinefigure}

In conclusion, a correct measure of the absolute He content 
on the basis of the R parameter requires the construction, for
any given cluster, of a synthetic CMD which properly reproduces 
the distribution of the stars along the HB and in turn a meaningful
evaluation of the $t_{HB}$. However, this approach is beyond the
scope of the present investigation.

\section{Conclusions}

We have measured the helium-sensitive R parameter in 26 Galactic globular
clusters imaged with WFPC2 on board the Hubble Space Telescope.
Our calculated R values are based on star counts that are corrected
for completeness and tested for radial variations within each cluster.
The high quality HST photometry also permits more clear separation
of the HB, AGB, and RGB stars.

The comparison between predicted and empirical R values appears to be
consistent with the absolute He abundance being lower than that found
from the observations of HII regions and from the primordial
nucleosynthesis models. One approach to overcome this discrepancy is
to adopt a \cdo nuclear cross section about a factor of two smaller
than current canonical values.  We note that HB lifetimes depend not
only on nuclear reaction rates, but also on the efficiency of mixing
processes and on the algorithms adopted for handling these physical
mechanisms.  In fact, as recently suggested by Cassisi et al. (2000),
current algorithms adopted for quenching the "breathing pulses"
introduce a $\approx 5 \%$ uncertainty on $t_{HB}$.  As a consequence,
the R parameter cannot presently be absolutely calibrated in terms of
a helium abundance.

The only trend in our data set is an unphysical trend toward higher
helium abundance in the clusters of lowest metallicity.  These
clusters tend to have blue horizontal branches, and we argue that
longer HB lifetimes in high temperature HB stars likely account for
this trend.  In fact, the global trend in R with metallicity is well
accounted for by changes in HB lifetime as a function of metal
abundance.

The trend in R values of the metal-rich globular clusters in our
sample is consistent with an upper limit of 2.5 for the helium to
metal enrichment ratio (\dydz).
The increased dispersion in R for the intermediate
metallicity clusters may be caused by the RGB bump fading below the HB
luminosity at [Fe/H]$\approx -1$, causing a drop in the calculated RGB
lifetime.  We conclude that these factors make the R values of low and
intermediate globular clusters less useful in constraining the helium
abundance.  Accurate photometric data for metal-rich globular
clusters, however, do place an interesting constraint on
\dydz .

The GGCs in the Galactic bulge are certainly a key target to accomplish 
these measurements. It is not a trivial effort to collect high quality 
data for such clusters, since they are often affected by high absolute 
and differential extinction (Ortolani et al. 1999), but the new NIR 
detectors should allow us to overcome these difficulties and to 
secure accurate data for a sizable fraction of cluster stars.  
A reconsideration of the bulge clusters and field population would
also be in order given the results of Minniti (1995); when his counts
are corrected for the contribution of AGB stars, the bulge fields
have R values of 1.7 to 2, higher than those of the most metal-rich
clusters in our sample. 

The promise of the R parameter being a constraint on the
primordial helium abundance remains frustratingly unfulfilled.
Two further thorny problems affect sound empirical estimates 
of this parameter, namely the foreground contamination and the 
radial population gradients. The former can be overcome by including 
only the innermost regions, whereas the latter is an open problem, 
since we still lack a systematic and quantitative estimate of this 
effect in GGCs from the very center to the tidal radius (Walker 1999).  
The precise comparison of synthetic H-R diagrams
with observational data may be required to achieve further progress.

\acknowledgements 
We thank an anonymous referee for his/her pertinent comments and 
suggestions that improved the content and the readability of the 
paper. This work was supported by MURST under the projects: "Stellar 
Dynamics and Stellar Evolution in Globular Clusters" ( G.\ P. \& M.\ Z.) 
and "Stellar Evolution" (G.\ B. \& S.\ C.).  S.G.D., and R.M.R.\
acknowledge support by NASA through grant GO-6095 and GO-7470 from the
Space Telescope Science Institute.  Partial support by ASI and CNAA is
also acknowledged.

\end{document}